\documentclass[12pt]{article}
\usepackage[english]{babel}
\usepackage{amsmath,amsthm}
\usepackage{amsfonts}
\usepackage{graphicx}
\usepackage{enumerate}
\usepackage[usenames,dvipsnames]{color}
\usepackage{subfigure}
\usepackage[right,pagewise,displaymath, mathlines]{lineno}
\usepackage{epstopdf}
\usepackage{color}

\numberwithin{equation}{section}
\numberwithin{figure}{section}
\numberwithin{table}{section}

\newtheorem{theorem}{Theorem}[section]
\newtheorem{corollary}[theorem]{Corollary}

\newtheorem{note}{Note}[section]

\numberwithin{equation}{section}

\begin{document}

To appear in \textit{The Mathematical Scientist} (Applied Probability Trust), 1st issue of 2017

\begin{center}
{\Large\bf Should we opt for the Black Friday discounted\\ price
 or wait until the Boxing Day?}

\vspace*{7mm}

\noindent {\large Jiang Wu$^{1,\,2}$ and Ri\v{c}ardas Zitikis$^{2}$}

\bigskip

$^{1}$\textit{School of Economics, Central University of Finance and Economics, Beijing 100081, P.R.~China.} E-mail: jwu447@uwo.ca
\medskip

$^{2}$\textit{Department of Statistical and Actuarial Sciences,
University of Western Ontario, London, Ontario N6A 5B7, Canada.} E-mail:  zitikis@stats.uwo.ca

\bigskip

\begin{quote}

\textbf{Abstract.} We derive an optimal strategy for minimizing the expected loss in two-period economy when a pivotal decision needs to be made during the first time-period and cannot be subsequently reversed. Our interest in the problem has been motivated by the classical shopper's dilemma during the Black Friday promotion period, and our solution crucially relies on the pioneering work of McDonnell and Abbott on the two-envelope paradox.

\medskip

\end{quote}

\end{center}

\section{Motivation}
\label{introduction}

When needing a laptop in the Fall of 2015, one of the authors of this article was looking for a good on-line deal. Benefiting from the advantages of on-line shopping, he acquired considerable information on the Lenovo T540 laptop price. Two forthcoming time periods were of immediate interest: the \textit{Black Friday} promotion from 27 November to 3 December 2015, and the \textit{Boxing Day} promotion from 26 December 2015 to 3 January 2016.

Obviously, it was prudent to wait until the Black Friday promotion period, which revealed the discounted price of 1,431.01 Canadian dollars for the laptop, but it was not obvious at that moment whether he wanted to buy the laptop at that price or wait until the Boxing Day promotion. (The price of Lenovo T540 laptop during the Boxing Day promotion period turned out to be 1,461.60 Canadian dollars, which was not, of course, known during the Black Friday promotion period.) Can there be a strategy for making good decisions during the Black Friday promotion period?

The present article aims at answering this question by deriving an optimal strategy that minimizes the expected buying price. The idea for tackling this problem stems from the pioneering work of McDonnell and Abbott (2009) on the two-envelope paradox, with further far-reaching considerations by McDonnell et al. (2011). It also relies on some of the techniques put forward by Egozcue et al. (2013) who have extended the aforementioned works beyond the two-envelope paradox. We next recall the paradox itself and in this way clarify its connection with our current problem. This will also provide useful hints on how we have resolved the problem and what challenges encountered.

Namely, there are two sealed and identically looking envelopes, with one containing twice more money than the other one. We randomly pick one envelope and open it. Let $x$ be the amount of money that we find. Should we keep the amount or swap the envelopes and have only what we find in the second envelope? The paradox is that if we swap, then with the probability $1/2$ we shall find either $x/2$ or $2x$ amount of money, and this gives us the average $(x/2)/2+(2x)/2=1.25x$, which is larger than $x$ that we have found in the first envelope. This argument suggests to always swap the envelopes.

For many years economists, engineers, mathematicians, statisticians, and others have worked on this paradox, with various competing and complementing solutions suggested. McDonnell and Abbott (2009) put the idea that the optimal strategy should be based on the information (i.e., the amount $x$) that we acquire after opening the first envelope, and then incorporating additional considerations in order to arrive at a threshold, say $x_0$, that would delineate those values of $x$ that would suggest swapping, or not swapping, the envelopes. This pioneering and inspiring solution of the paradox by McDonnell and Abbott (2009), with further refinements by McDonnell et al. (2011), has been extensively discussed in the scientific and popular literature.

In the present paper we demonstrate how this solution can be adjusted and extended to solve other problems, spanning well beyond the two-envelope paradox. In particular, just like McDonnell and Abbott (2009), we also derive a threshold-based strategy for accepting or rejecting the price offered during the first time-period. Challenges naturally arise, including the arbitrariness of prices, which are not the aforementioned $x$, $x/2$ or $2x$ anymore. The need for incorporating elements of behavioural economics and rational decision-making arise. These and other considerations inevitably introduce additional mathematical and probabilistic challenges, which we shall discuss in great detail below, and within the context of the shopper's dilemma noted earlier.

The rest of this paper is organized as follows. In Section \ref{results}, which contains our main results, we first lay out the necessary mathematical background and then derive two optimal strategies: one when there is no guessing of the second time-period price, and the other when such a guessing can take place. Section \ref{conclusions} concludes the main part of the paper with a brief overview of our main contributions. In technical Appendices \ref{app-a} and \ref{app-b}, we discuss price modeling and parameter specifications of practical relevance. Proofs of the main results are given in Appendix~\ref{app-c}.

\section{Main results}
\label{results}

We start out by carefully describing the decision-making process, and also introduce the necessary notation. First, the prospective buyer contacts a salesperson during the first time-period. To somewhat simplify the problem, we assume that there are two kinds of salespersons:
\begin{enumerate}
\item[i)]
those, call them $L$, who tend to offer larger discounts and thus \textit{lower} prices $X_{L}$;
\item[ii)]
others, say $H$, who tend to offer smaller discounts and thus \textit{higher} prices $X_{H}$.
\end{enumerate}
Both $X_{L}$ and $X_{H}$ are random variables. We denote their joint cumulative distribution function (cdf) by $F_{X_L,X_H}(x,y)$ and assume that it is absolutely continuous, that is, has a density $f_{X_L,X_H}(x,y)$, which is a natural assumption in the current context.

Let $\Pi_{1}$ denote the random variable that takes the values $L$ and $H$  depending on which (kind of) salesperson takes the prospective buyer's call during the first time-period. Hence, when $\Pi_{1}=L$, then the price is $ X_{L} $, and when $\Pi_{1}=H$, then $X_{H}$. Note that for the management and the salespersons, the actual prices might be pre-determined and thus known, but for the buyer they are unknown and thus treated as random variables  $X_{L}$ and $X_{H}$ following some distributions, with an outcome of one of the random variables observed during the first time-period.

Once the prospective buyer learns the price during the first time-period (i.e., Black Friday), he has two options: to either accept the offer or reject it and then inevitably wait till the second time-period (i.e., Boxing Day). If the buyer thinks that the first time-period offer is good enough, he accepts it and the purchasing process ends, but if the buyer rejects the offer, then he has to wait until the second time-period and then inevitably accept whatever offer is made to him at that time, because he needs a laptop and the regular price is less attractive than any of the discounted ones.

Let $\Delta_{1}$ denote the random variable that represents the prospective buyer's decision during the first time-period: $\Delta_{1}=A$ if the buyer accepts the first-period offer and $\Delta_{1}=R$ if he rejects it. The aim of the present article is to offer an optimal strategy that minimizes the expected value $\mathbf{E} [X]$ of the buying price $X$, which could be either $X_{L}$ or $X_{H}$ depending the the buyer's decision during the first time-period.

It is natural to assume -- unless the buyer possesses insider's information but we do not consider this case in the paper -- that the buyer does not know who, $L$ or $H$, is making offers during the first time-period. Hence, the buyer's decision $\Delta_{1}=A$ to accept the price offered during the first time-period does not depend on who, $L$ or $H$, makes the offer -- it depends only on the price, $X_{\Pi_{1}}$ or $X_{\Pi_{2}}$, being offered, where $\Pi_{1},\Pi_{2}\in \{ L, H\}$ and $\Pi_{1}\neq \Pi_{2}$. In rigorous probabilistic terms, this means that the probability $\mathbf{P}[\Delta_{1}=A \mid X_{\Pi_{1}},X_{\Pi_{2}}, \Pi_{1}]$ is equal to $\mathbf{P}[\Delta_{1}=A
\mid X_{\Pi_{1}},X_{\Pi_{2}}]$ for all possible outcomes of the prices $X_{\Pi_{1}}$ and $X_{\Pi_{2}}$, and for every salesperson $\Pi_{1}\in \{ L, H\}$. We note in passing that the just noted probability $\mathbf{P}[\Delta_{1}=A \mid X_{\Pi_{1}},X_{\Pi_{2}}]$ will later define a certain strategy function (equation (\ref{strategy-s}) in Appendix \ref{app-c}) which will play a crucial role in deriving actionable strategies that we spell out in Theorems \ref{th-1a} and \ref{th-1b} below. One may rightly argue further and suggest that in the context of our motivating problem, the probability of the decision $\Delta_{1}=A$ does not depend on the second price  $X_{\Pi_{2}}$, and we shall indeed tackle this case most prominently (Theorem \ref{th-1a}; also equation (\ref{strategy-s2}) in Appendix \ref{app-c})

Finally before formulating Theorem \ref{th-1a}, we introduce yet another quantity, $p(x,y)$, which plays a pivotal role throughout the rest of the paper. Namely, conditionally on the prices $X_L$ and $X_H$, let $p(X_L,X_H)$ be the probability that the salesperson $L$ is in charge of making offers during the first time-period. Naturally, the company's management knows who, $L$ or $H$, is making offers during the first time-period, but the decision making process that we are concerned with is from the perspective of the buyer, who can only use her/his intuition or some educated arguments such as those offered by economic theories in order to guess who, $L$ or $H$, could possibly be, and with what likelihood, in charge of making offers during the first time-period. Hence, $p(X_L,X_H)=\mathbf{P}(\Pi_{1}=L\mid X_{L},X_{H})$ or, in other words, $p(x,y)=\mathbf{P}(\Pi_{1}=L\mid X_{L}=x,X_{H}=y)$ for all possible outcomes $x$ and $y$ of the random prices $X_L$ and $X_H$, respectively. We shall discuss the probability $p(x,y)$ in great detail in Appendix \ref{app-b}, including its modelling and accompanying economic considerations.

\begin{theorem}\label{th-1a}
When the price of the first time-period is $v$ and there is no attempt to guess the possible price to be offered during the second time-period, then  the strategy that minimizes the expected buying price is to accept the offer when $h(v)\le 0$ and to reject it when $h(v)> 0$, where the ``no guessing strategy'' function $h(v)$ is
\begin{equation}\label{h-v}
h(v)
=\int_0^{\infty} (v- w)p(v,w)f_{X_L,X_H}(v,w)d w
+\int_0^{\infty} (v- w)\big (1-p(w,v)\big )
f_{X_L,X_H}(w,v)d w .
\end{equation}
\end{theorem}

\begin{figure}[h!]
\centering
\includegraphics[height=5.5cm,width=12cm]{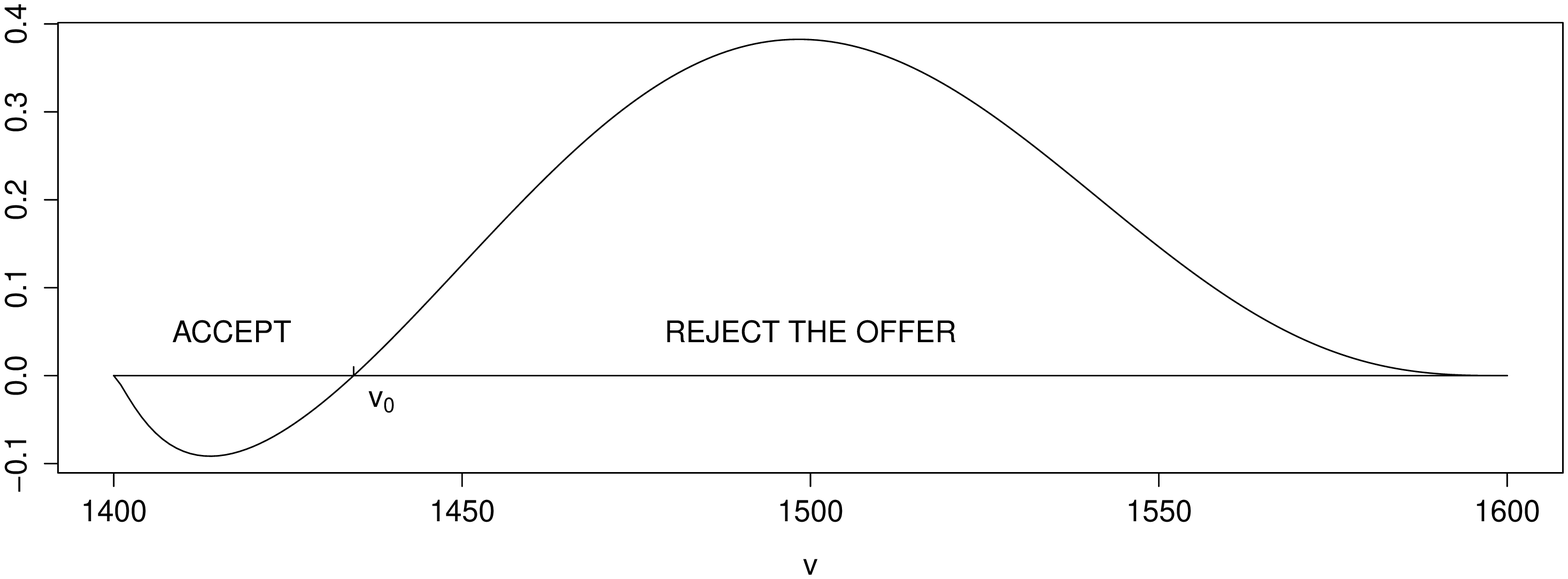}
\caption{The no-guessing-strategy function $h(v)$.}
\label{hv}
\end{figure}

We have visualized the strategy function $h(v)$ in Figure \ref{hv}
with its properties, modelling, and specific parameter choices to be discussed next. To begin with, it is natural to think of the joint density $f_{X_L,X_H}(v,w)$ as a continuous function with compact support $[x_{\min},x_{\max}]\times [x_{\min},x_{\max}]$, where $x_{\min}$ is the reservation price from the supply side (i.e., computer technology company) and $x_{\max}$ is the reservation price from the demand side (i.e., consumer). Hence, $f_{X_L,X_H}(w,v)=0$ when $v$ or $w$, or both, are outside the interval $(x_{\min},x_{\max})$.

We also expect that under normal circumstances there should be a point $v_0\in (x_{\min},x_{\max})$ such that $h(v)<0$ (accept the offer) for all $v\in (x_{\min},v_0)$ and $h(v)>0$ (reject the offer) for all $v\in (v_0,x_{\max})$, with $h(v_0)=0$. We indeed see this pattern in our illustrative Figure \ref{hv}, where and elsewhere when graphing in this paper we set the reservation prices to $x_{\min}=1,400$ and $x_{\max}=1,600$ Canadian dollars. For other specifications, including the underlying economic theories, modelling, and choices of $f_{X_L,X_H}(v,w)$ and $p(v,w)$, we refer to Appendices \ref{app-a} and \ref{app-b}.

We note that under the aforementioned specifications, the point where the function $h(v)$ crosses the horizontal axis is $v_0\approx 1,434.43$, which delineates the acceptance (to the left) and rejection (to the right) regions. It should also be noted that the inclusion of the point $v_0$ into the acceptance region is arbitrary: whenever $v\in (x_{\min},x_{\max})$ is such that $h(v)=0$, we could very well flip a coin to decide whether to accept the offer or reject it and wait until the next promotion period. In general, the form of the function is of interest because its lows and highs tell us how confident we can be when making decisions (accept or reject) depending on the price $v$ and the likelihood of this price being offered.

\begin{theorem}\label{th-1b}
When the price of the first time-period is $v$ and the guessed price to be offered during the second time-period is $w$, then the strategy that minimizes the expected buying price is to accept the offer when $h(v,w)\le 0$ and to reject it when $h(v,w)> 0$, where the ``guessing strategy'' surface is
\begin{equation}\label{h-vw}
h(v,w)=(v- w)p(v,w)f_{X_L,X_H}(v,w)
\\
+(v- w)\big (1-p(w,v)\big )f_{X_L,X_H}(w,v).
\end{equation}
\end{theorem}

\begin{figure}[h!]
\centering
\includegraphics[height=6cm,width=7cm]{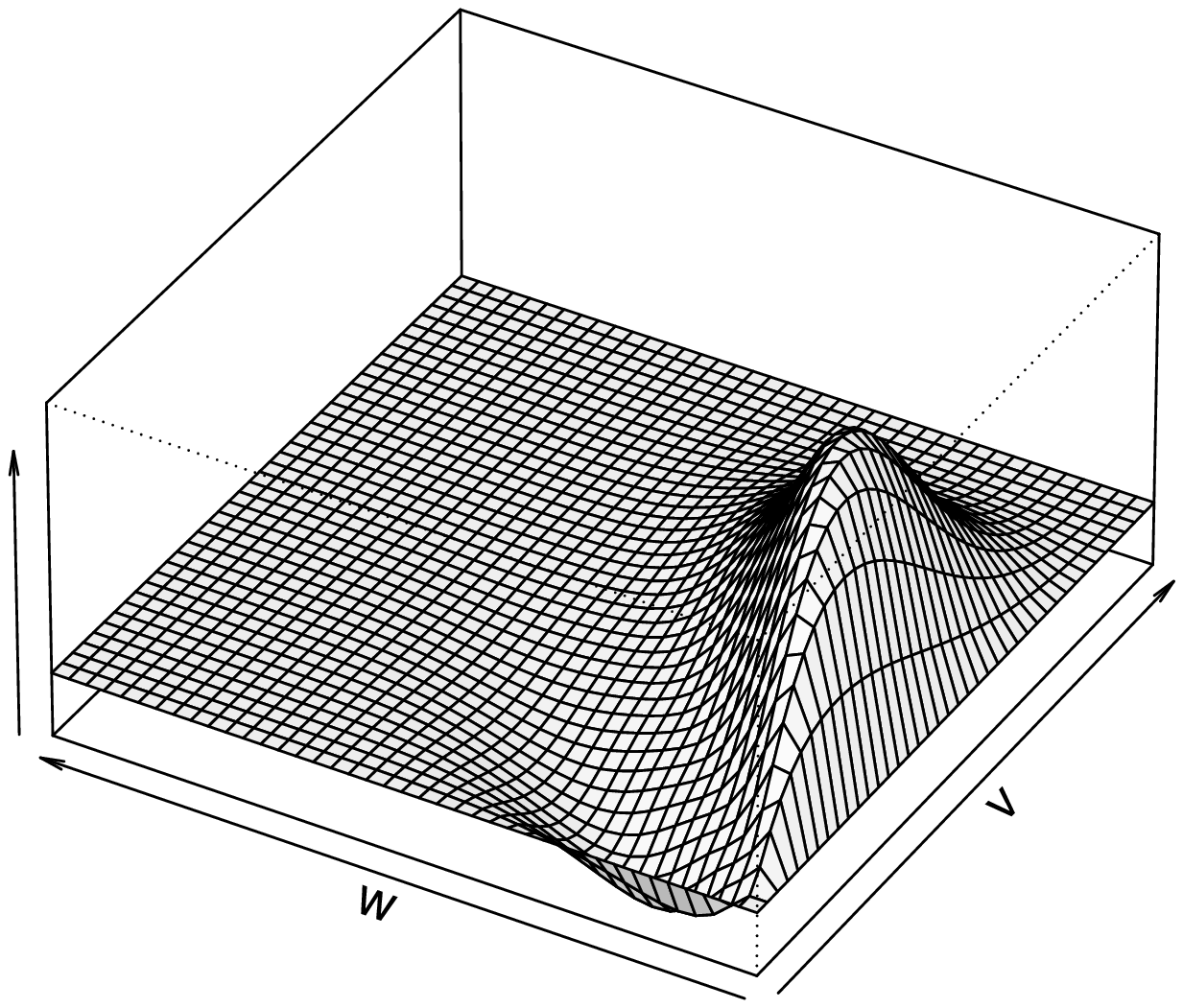}
\includegraphics[height=6.5cm,width=6cm]{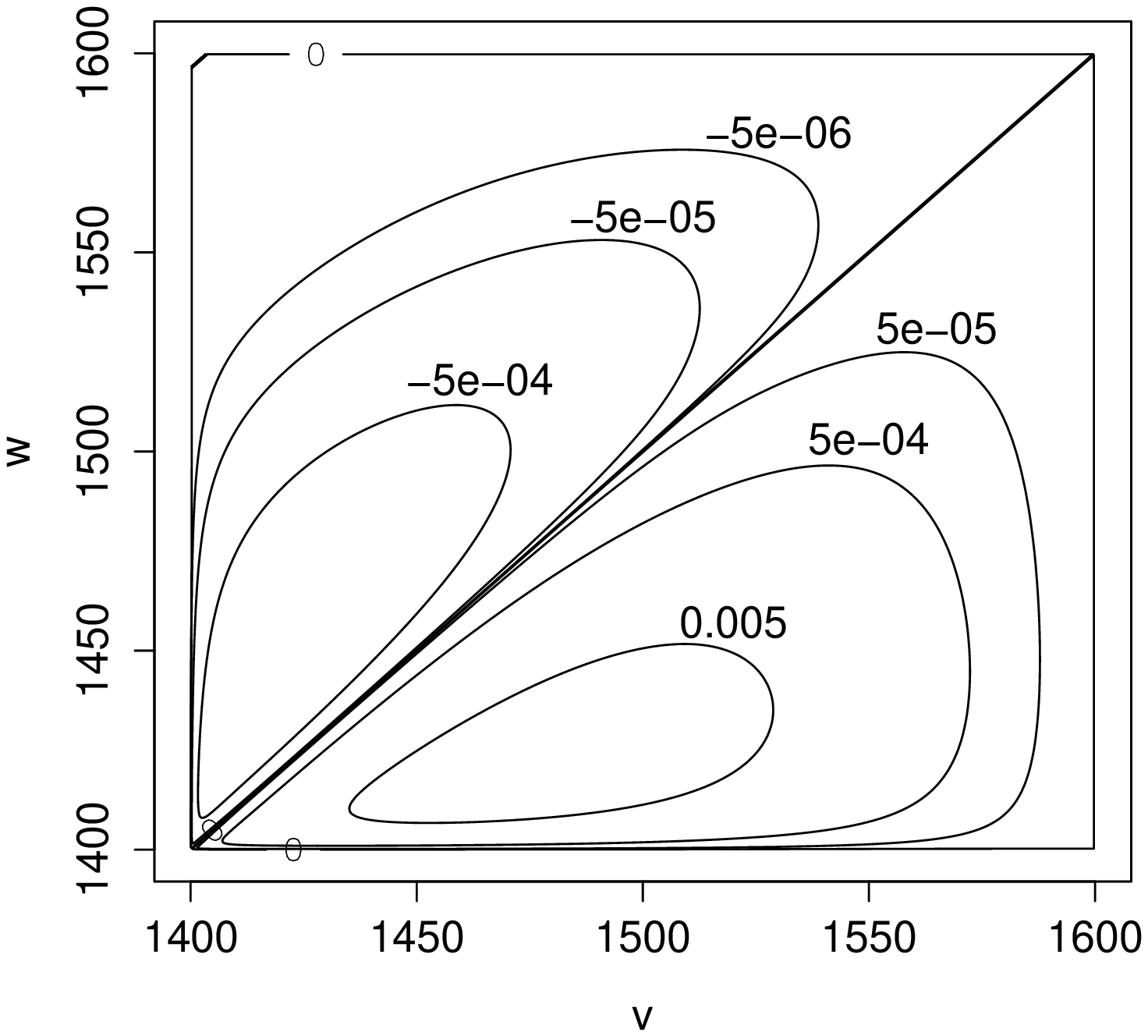}
\caption{The guessing-strategy surface $h(v,w)$ (left) and its contours (right).}
\label{hvw-contours}
\end{figure}

Note that $h(v,w)=0$ when $v=w$, which is natural. It is also natural to expect that $h(v,w)<0$ (i.e., accept the first time-period price $v$) whenever $v<w$, and $h(v,w)>0$ (i.e., reject the first time-period price $v$) whenever $v>w$. These features are of course clearly seen from formula (\ref{h-vw}) because the functions $f_{X_L,X_H}(v,w)$ and $p(v,w)$ are positive in the interior  $(x_{\min},x_{\max})\times (x_{\min},x_{\max})$ of their supports. We also see these features in Figure \ref{hvw-contours}
where we have depicted the surface $h(v,w)$ and its contours in the case of the functions $f_{X_L,X_H}(v,w)$ and $p(v,w)$ specified in Appendices \ref{app-a} and \ref{app-b}.  The lows and highs of the surface tell us how confident we can be when making decisions (accept or reject) depending on the prices $v$ and $w$, as well as on the likelihood of these prices being offered.

We finish this section with a brief discussion of possible models for $p(v,w)$, with further details provided in Appendix \ref{app-b} below. Namely, upon recalling that $p(v,w)$ is the  probability that $L$ makes an offer during the first time-period, given that the offers of $L$ and $H$ are $v$ and $w$ respectively, it is natural to model $p(v,w)$ as $F(v-w)$ with some cdf $F$ such that $F(0)=1/2$. In Appendix \ref{app-b}, for example, we shall use the beta cdf with identical shape parameters, in which case we have the equation $F(x)=1-F(-x)$ for all $x$ and thus, in particular, the requirement $F(0)=1/2$. The following corollary to Theorems  \ref{th-1a} and \ref{th-1b} deals with this special case.

\begin{corollary}\label{cor-1}
Let $p(v,w)=1-p(w,v)$ for all $v$ and $w$. Then the guessing-strategy surface is
\[
h(v,w)=(v- w)p(v,w)
\big ( f_{X_L,X_H}(v,w)+f_{X_L,X_H}(w,v)\big ),
\]
and the no-guessing-strategy function is
\[
h(v)
=\int_0^{\infty} h(v,w) d w.
\]
\end{corollary}

The earlier drawn Figures \ref{hv} and \ref{hvw-contours} are based on Corollary \ref{cor-1} with practically relevant modelling of the functions $f_{X_L,X_H}(v,w)$ and $p(v,w)$ discussed in Appendices \ref{app-a} and \ref{app-b}. For the definition of $f_{X_L,X_H}(v,w)$, we refer to equation (\ref{ex-f}), and for that of $p(v,w)$ to equation (\ref{ex-p}). Since closed-form formulas are unwieldy, we therefore employed numerical integration techniques, which proved to be fast and efficient and thus appealing from the practical point of view.

\section{Conclusions}
\label{conclusions}

Decision making in two-period economy has been a topic of much interest for researchers in various fields, including economics, engineering, and finance. Mathematicians, philosophers, and statisticians have also contributed significantly to the area. In the present paper we have tackled this topic in the setup that concerns shoppers facing two price-discount periods and needing a strategy for making beneficial (for them) decisions. Guided by economic theories and rigorous probabilistic considerations, we have developed a practically sound model for making such decisions. In particular, we have shown how to derive, analyze, and use strategy functions, which not only delineate the acceptance and rejection regions for the first-come offers but also tell us how confident we can be when making such decisions.

\appendix

\section{Background price models and the choice of $f_{X_L,X_H}(x,y)$}
\label{app-a}

If the salespersons $L$ and $H$ were in total isolation, their offered prices would be outcomes of two independent random variables, which we denote by $X^0_L$ and $X^0_H$, both taking values in the interval $ [x_{\min},x_{\max}]$. It is natural to assume, for example, that $X^0_L $ and $X^0_H $ are beta distributed on $[x_{\min},x_{\max}]$ with positive shape parameters $(\alpha_L,\beta_L)$ and $(\alpha_H,\beta_H)$, respectively, that is,
\[
f_{X^0_L}(x)={(x-x_{\min})^{\alpha_L-1}
(x_{\max}-x)^{\beta_L-1}\over B(\alpha_L,\beta_L)\rho^{\alpha_L+\beta_L-1}}, \quad x_{\min}<x< x_{\max},
\]
and
\[
f_{X^0_H}(x)={(x-x_{\min})^{\alpha_H-1}
(x_{\max}-x)^{\beta_H-1}\over B(\alpha_H,\beta_H)\rho^{\alpha_H+\beta_H-1}}, \quad x_{\min}< x< x_{\max} ,
\]
where
\[
\rho=x_{\max}-x_{\min}
\]
is the range of possible prices. Given the earlier noted numerical values of $x_{\min}$ and $x_{\max}$, we have $\rho=200$. When graphing throughout this paper, we always set the parameter values to $(\alpha_L,\beta_L)=(2.5,4.5)$ and $(\alpha_H,\beta_H)=(4.5,1.5)$.

The observable prices offered by $L$ and $H$ are, however, not $X^0_L$ and $X^0_H$ but those that have been influenced by, e.g., the company's marketing team or management. This naturally leads us to the background price model, called the background risk model in the economic and insurance literature, which we choose to be multiplicative (cf., e.g., Franke et al.~2006, 2011; Asimit et al.~2016; and references therein).

Namely, suppose that $Y\in [x_{\min},x_{\max}]$ is the (random) price that the company's management would think appropriate, and which therefore influences the actual decisions of $L$ and $H$. The multiplicative background model would suggest that the observable prices $X_L $ and $X_H $ are of the form $X_L Y$ and $X_H Y$, but when defined in this way they are outside the natural price-range $[x_{\min},x_{\max}]$. To rectify the situation, we first standardize the prices $X^0_L$, $X^0_H$, and $Y$ using the equations $Z^0_L = (X^0_L-x_{\min})/\rho $, $Z^0_H = (X^0_H-x_{\min})/\rho $, and $Z = (Y-x_{\min})/\rho $, and then model the observable prices as
$X_L=Z^0_L Z \rho +x_{\min}$ and $X_H=Z^0_H Z\rho +x_{\min}$. Throughout the rest of this section, we assume that $X^0_L $, $X^0_H $ and $Y$ are independent and thus, in turn, their standardized versions $Z^0_L $, $Z^0_H $ and $Z$ are such as well. We shall soon find this assumption convenient; in fact, it is a natural assumption.

\begin{figure}[h!]
\centering
\subfigure[Pdf's of $X^0_L$ (dotted) and $X_L$ (solid).]
{\includegraphics[height=5.5cm,width=6cm]{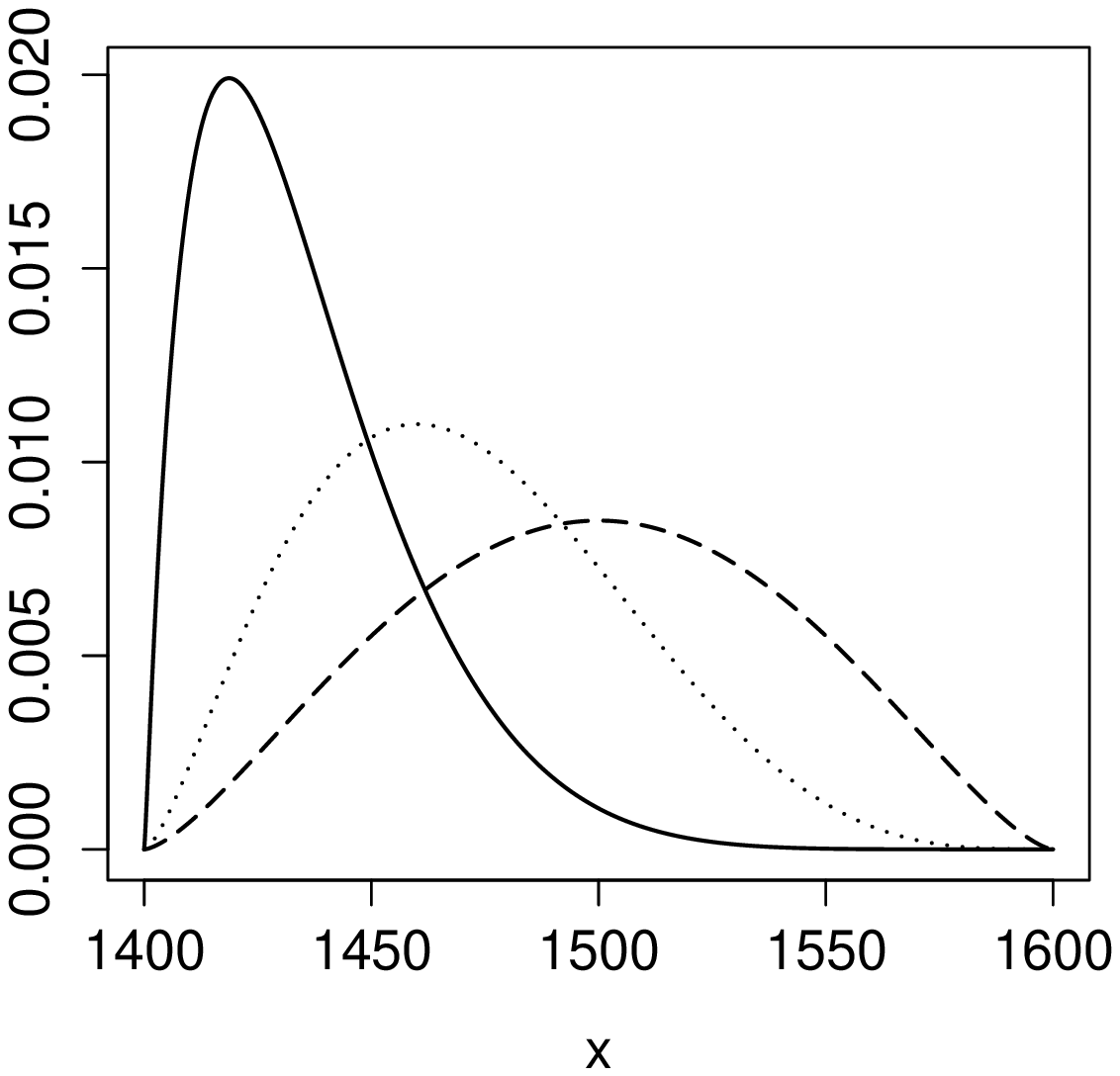}}
\subfigure[Pdf's of $X^0_H$ (dotted) and $X_H$ (solid).]
{\includegraphics[height=5.5cm,width=6cm]{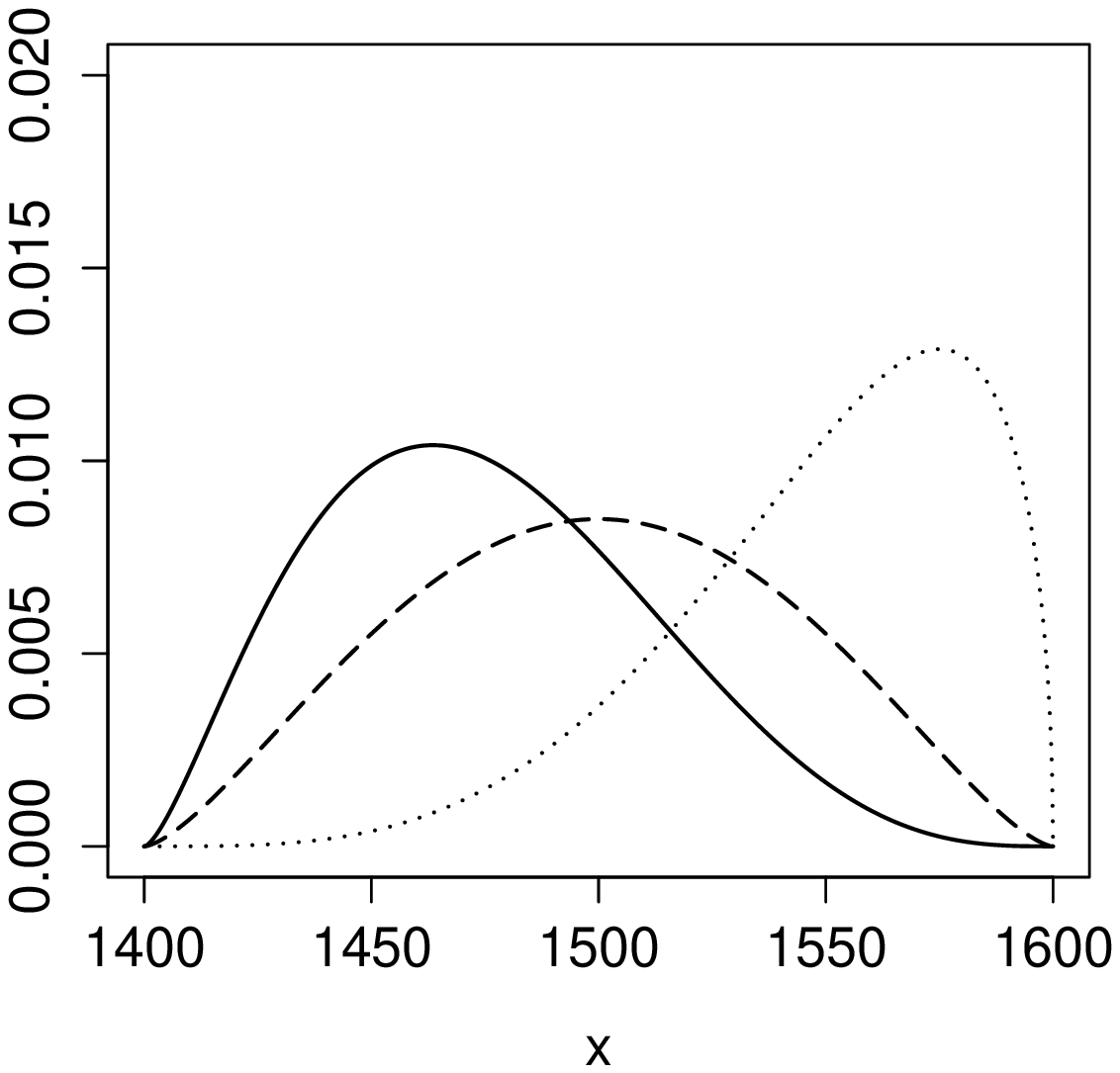}}
\caption{The influence of $Y$ (dashed pdf in both panels) on $X^0_L$ and $X^0_H$.}
\label{fig-1}
\end{figure}

Let, for example, $Y$ follow the beta distribution on the interval $ [x_{\min},x_{\max}]$ with some (positive) shape parameters $(\alpha_0,\beta_0)$. Then the pdf $f_{Z}(t)$ of $Z$ is equal to
\[
f_{\alpha_0, \beta_0}(t):={t^{\alpha_0-1}
(1-t)^{\beta_0-1}\over B(\alpha_0,\beta_0)}, \quad 0< t < 1.
\]
The marginal pdf's of $X_{L}$ and $X_{H}$ are
\begin{equation}
f_{X_L}(x)
= { 1 \over \rho  }\int_{t_x}^{1} f_{\alpha_L, \beta_L}\bigg ({x-x_{\min}\over t\rho }\bigg ) {1\over t} f_{\alpha_0, \beta_0}(t) dt
\label{marg-l}
\end{equation}
and
\begin{equation}\label{marg-h}
f_{X_H}(x)={ 1 \over \rho  }\int_{t_x}^{1} f_{\alpha_H, \beta_H}\bigg ({x-x_{\min}\over t\rho }\bigg ) {1\over t} f_{\alpha_0, \beta_0}(t) dt,
\end{equation}
respectively, where $t_x= (x-x_{\min})/\rho $. We have depicted the pdf's in Figure \ref{fig-1}
using $(\alpha_0,\beta_0)=(2.5,2.5)$ and the earlier noted parameter choices for $(\alpha_L,\beta_L)$ and $(\alpha_H,\beta_H)$.

\begin{note}\rm
Pdf's (\ref{marg-l}) and (\ref{marg-h}) can be written in terms of the hypergeometric function ${_{2}}F_1 $ as follows:
\begin{multline*}
f_{X_L}(x)
= { 1 \over \rho }
{\Gamma(\alpha_L+\beta_L)\Gamma(\alpha_0+\beta_0) \over
\Gamma(\beta_L+\beta_0)\Gamma(\alpha_L)\Gamma(\alpha_0)}
t_x^{\alpha_L-1} (1-t_x)^{\beta_L+\beta_0-1}
\\
\times
{_{2}}F_1 (\beta_0,\alpha_L+\beta_L-\alpha_0; \beta_L+\beta_0; 1-t_x),
\end{multline*}
and an analogous expression holds for $f_{X_H}(x)$. These expressions follow from well-known formulas for the pdf of the product of two independent type-I beta random variables (Nagar and Zarrazola, 2004; also Nagar et al., 2014; and references therein).
\end{note}

\begin{figure}[h!]
\centering
\includegraphics[height=6cm,width=7cm]{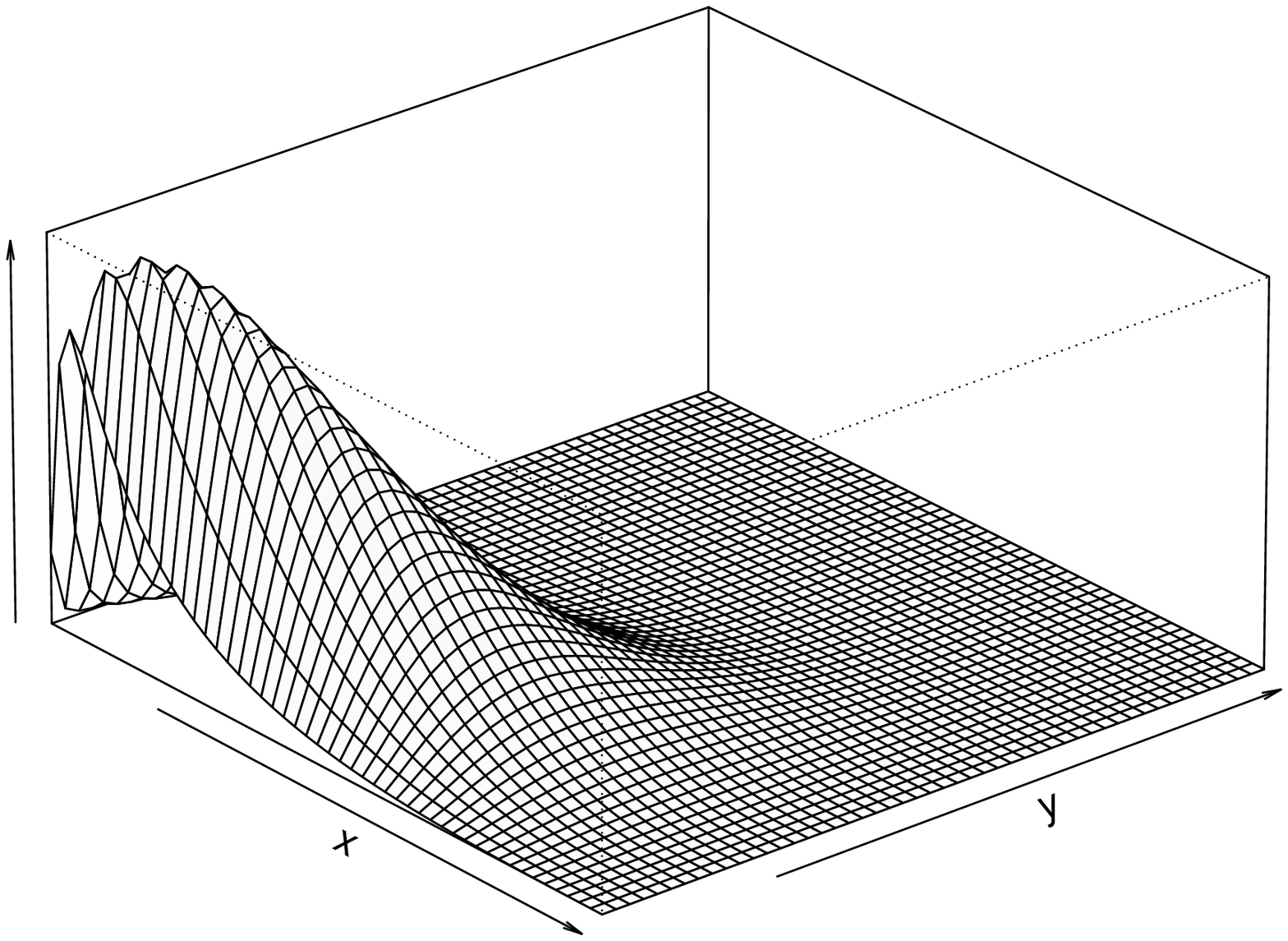}
\includegraphics[height=6cm,width=5.5cm]{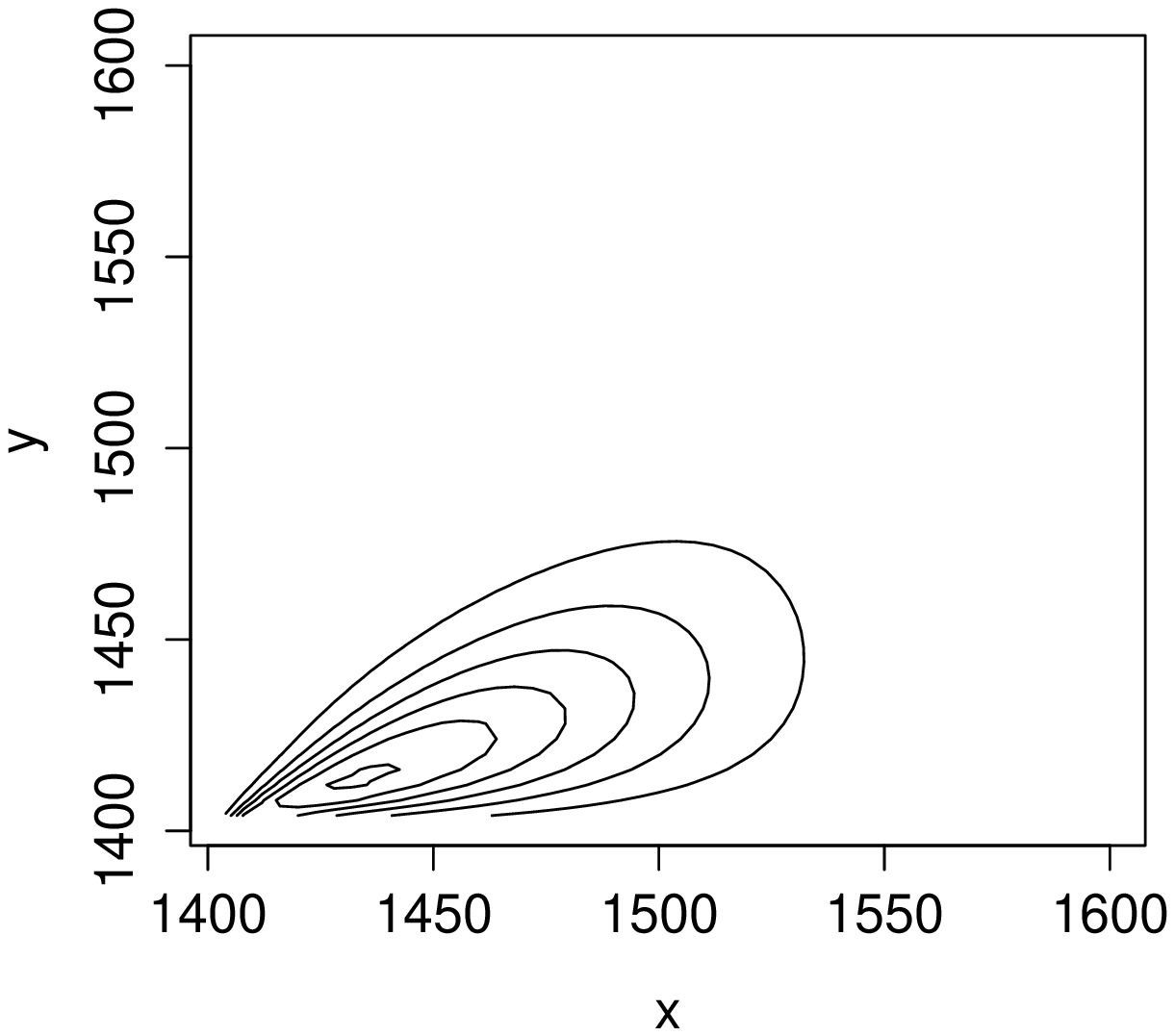}
\caption{The joint pdf of $X_{L}$ and $X_{H}$ (left) and its contours (right).}
\label{fig-2}
\end{figure}

Similarly to equations (\ref{marg-l}) and (\ref{marg-h}), we derive the joint pdf
\begin{equation}\label{ex-f}
f_{X_L,X_H}(x,y)
={ 1 \over \rho ^2 }\int_{t_{x,y}}^{1} f_{\alpha_L, \beta_L}\bigg ({x-x_{\min}\over t\rho }\bigg ) f_{\alpha_H, \beta_H}\bigg ({y-x_{\min}\over t\rho }\bigg ) {1\over t^2} f_{Z}(t) dt,
\end{equation}
where $t_{x,y}=(\max\{x,y\}-x_{\min})/\rho $.
We have depicted this pdf in Figure \ref{fig-2}.

\section{Economic justification and modelling of $p(x,y)$}
\label{app-b}

To begin with, we view our problem within the context of two-period (also known as two-stage) economy. Indeed, it is reasonable to assume that many consumers who buy laptops during the first time-period have stronger buying intentions than most of the consumers who buy laptops during the second time-period. That is, the demand curves during the two periods have different demand elasticities of price: during the first time-period, the demand elasticity of price is smaller than that during the second time-period. Considering this difference, the buyers during the two periods can be viewed as two separate markets with different demand curves and different demand elasticities, and monopolistic firms would tend to enforce the division of selling prices during the two periods.

This leads us to the topic of third-degree price discrimination  (e.g., Schwartz, 1990; Aguirre et al., 2010), which in the monopolistic competitive market means that the same provider would charge different prices for similar goods or services in different consumer groups having different demand curves and demand elasticities, such as those who buy laptops during the first time-period and those who buy during the second period. In general, product differentiation is one of the key factors why firms have some degree of control over the prices, and the more successful a firm is at differentiating its products from other firms selling similar products, the more monopoly power the firm has. Product differences arise due to quality, functional features, design, and so on, and so there is imperfect substitution between products even when they are in the same category (e.g., Krugman, 1980; Head and Ries, 2001).

Hence, in the case of our problem concerning laptops, there is arguably a tendency to offer higher prices during the first time-period because the demand curve in the first period is a relatively inelastic demand curve, due to stronger buying intention. Hence, it is natural that firms would take advantage of this buying intention and offer higher prices during the first time-period. Consequently, it would seem that the higher the price is offered by $L$, the higher the probability that the consumer will receive an offer from $L$ during the first time-period. The higher the laptop price is offered by $H$, the lower the probability that an offer will come from $L$ during the first time-period. When $L$ and $H$ offer laptops at the same or similar price, then the probability of getting an offer from $L$ would be more or less the same as the probability of getting an offer from $H$. In view of these arguments, the following properties seem natural:
\begin{enumerate}
  \item[1)] $p(x,y)$ is non-decreasing in $x$, for every fixed $y\in (x_{\min},x_{\max})$;
  \item[2)] $p(x,y)$ is non-increasing in $y$, for every fixed $x\in (x_{\min},x_{\max})$;
  \item[3)] $p(x,y)=1/2$ whenever $x=y$.
\end{enumerate}

\begin{figure}[h!]
\centering
\includegraphics[height=6cm,width=7cm]{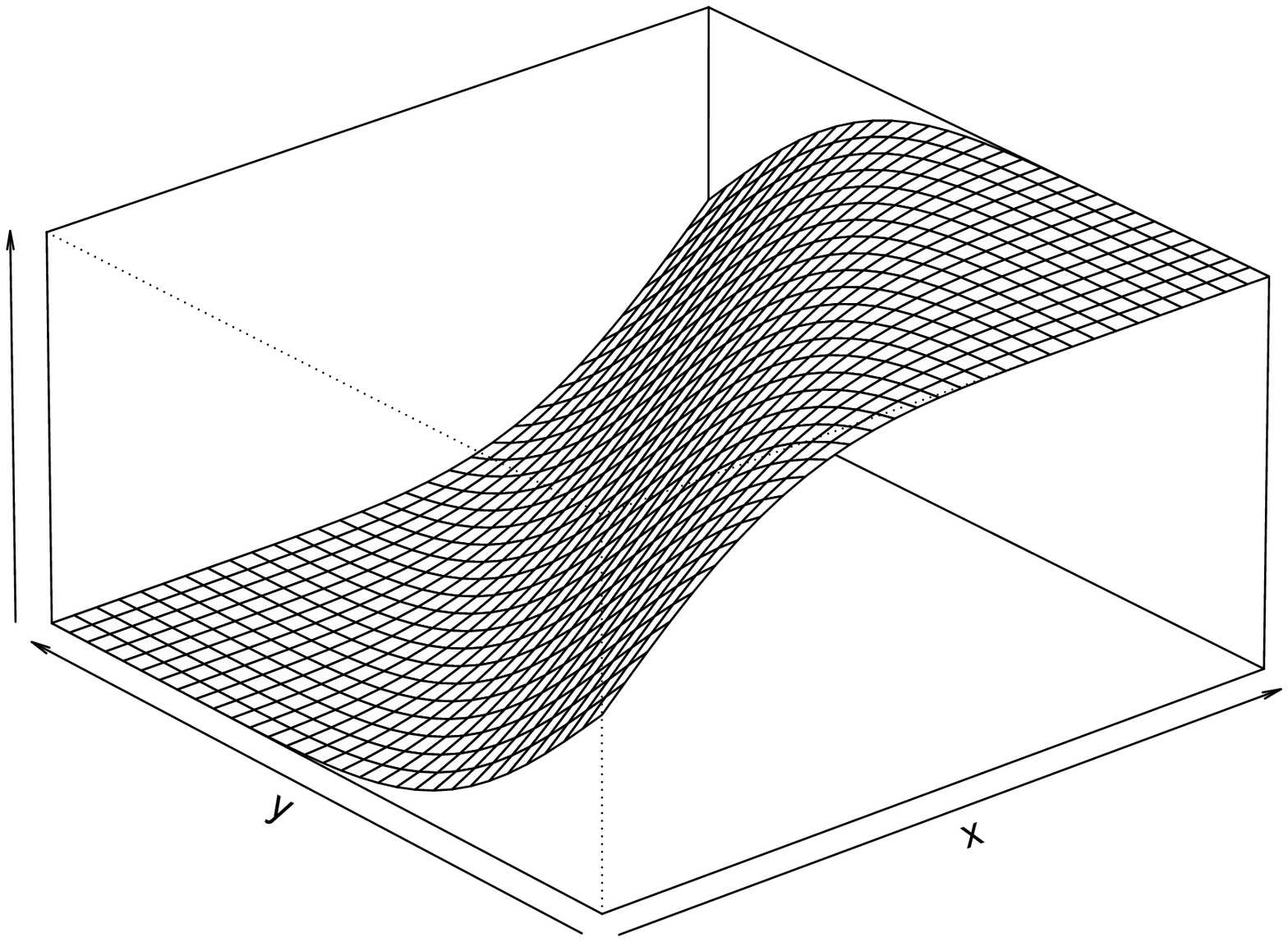}
\includegraphics[height=6cm,width=5.5cm]{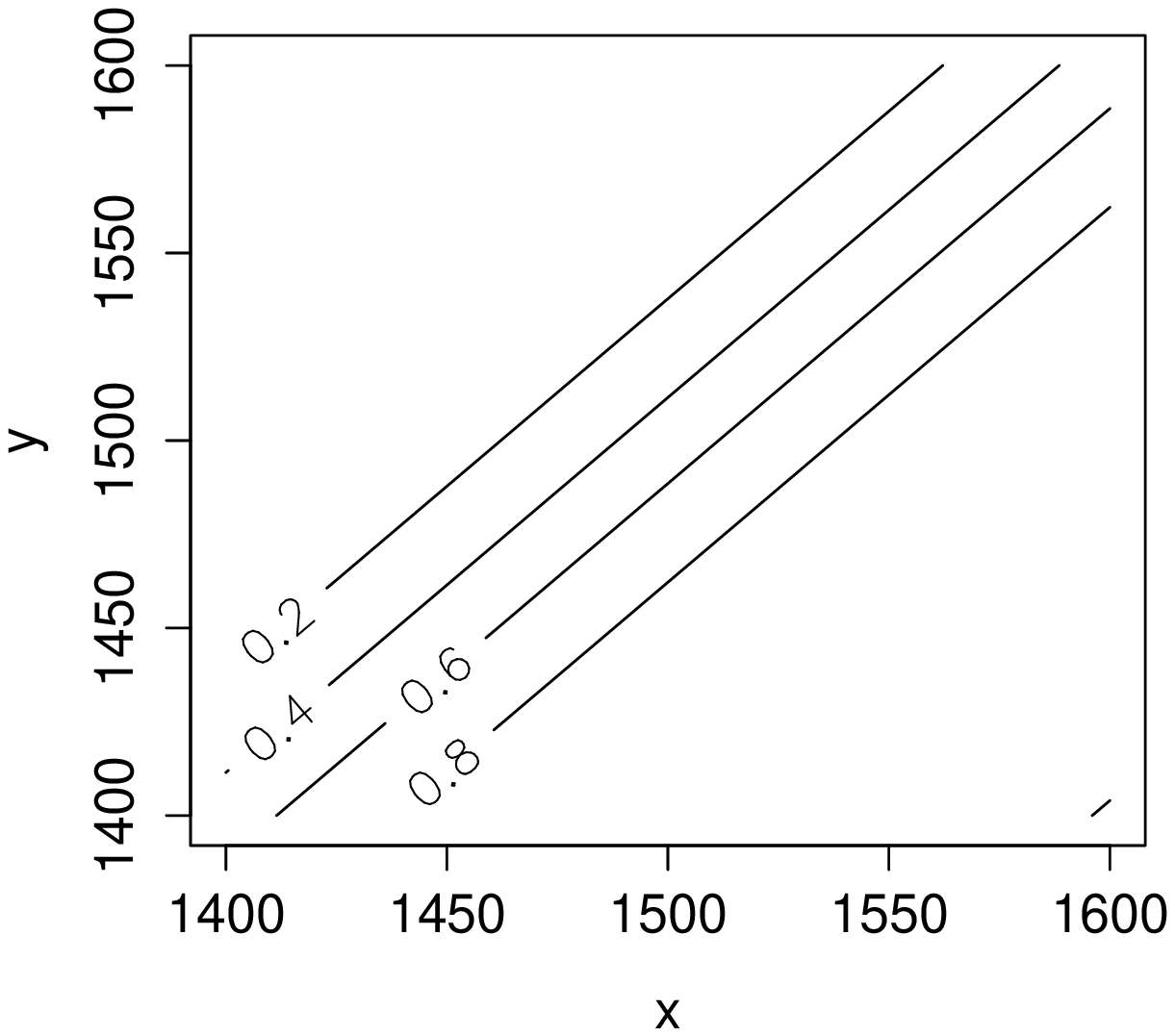}
\caption{The probability surface $p(x,y)$ (left) and its contours (right).}
\label{fig-3}
\end{figure}

We next suggest an example of $p(x,y)$ by setting it to be $F(x-y)$, where $F$ can be any cdf such that $F(0)=1/2$. For example, let $F$ be the beta cdf on the interval $[-\rho, \rho]$ with $\rho=x_{\max}-x_{\min}$ and equal shape parameters, say $\gamma >0$. That is,
\begin{equation}\label{ex-p}
  p(x,y)={1\over B(\gamma,\gamma)(2\rho)^{2\gamma-1}}\int_{-\rho}^{x-y} (\rho+t)^{\gamma-1}(\rho-t)^{\gamma-1}dt ,
\end{equation}
depicted in Figure \ref{fig-3} with the parameter $\gamma =10$, which we always use when graphing.

\section{Proofs of Theorems \ref{th-1a} and \ref{th-1b}}
\label{app-c}

Our goal is to derive a strategy that leads to the minimal expected value $\mathbf{E} [X]$ of the buying price $X$, which could be either $X_{L}$ or $X_{H}$ depending on the outcomes of the random variables $\Pi_{1}\in \{L,H\}$ and $\Delta_{1}\in \{A,R\}$. We start with the equation
\begin{equation}
\label{conditional-expectation-first}
\mathbf{E} [X]=\iint\mathbf{E} [X\mid X_{L}=x,X_{H}=y]dF_{X_L,X_H}(x,y)
\end{equation}
and then work with the conditional expectation inside the integral. Since the prices are non-negative, all the integrals throughout the proofs are from $0$ to $\infty$.

Given $X_{L}=x$ and $X_{H}=y$, the random variable $X$ can take only the values $x$ or $y$. Consequently, we have the equation
\begin{multline}\label{expectation-first}
\mathbf{E}[X\mid{X_{L}=x,X_{H}=y}]=x\mathbf{P}[X=x\mid{X_{L}=x,X_{H}=y}]
\\
+y\mathbf{P}[X=y\mid X_{L}=x,X_{H}=y].
\end{multline}
We next calculate the two probabilities on the right-hand of equation
\eqref{expectation-first} based on which of the two salespersons, $L$ or $H$, is making offers during the first time-period, and also on the consumer behavior during this period, who can either accept or reject the first-come offer.

To begin with, we employ the random variable $\Pi_{1}\in \{L,H\}$ and have
\begin{align}\label{rhandfir}
&\mathbf{P}[X =x\mid{X_{L}=x,X_{H}=y}]
\notag
\\
&=\mathbf{P}[X=x\mid{X_{L}=x,X_{H}=y,\Pi_{1}=L}]
\mathbf{P}[\Pi_{1}=L\mid{X_{L}=x,X_{H}=y}]
\notag
\\
&\quad +\mathbf{P}[X=x\mid X_{L}=x,X_{H}=y,\Pi_{1}=H]
\mathbf{P}[\Pi_{1}=H\mid{X_{L}=x,X_{H}=y}].
\end{align}
We next tackle the four probabilities on the right-hand side of equation \eqref{rhandfir}, starting with the first probability.

Using the random variable $\Delta_{1}\in \{A,R\}$, we have the equation
\begin{align}\label{rhandfirone}
&\mathbf{P}[X=x\mid{X_{L}=x,X_{H}=y,\Pi_{1}=L}]
\notag
\\
&=\mathbf{P}[X=x\mid{X_{L}=x,X_{H}=y,\Pi_{1}=L,\Delta_{1}=A}]
\mathbf{P}[\Delta_{1}=A\mid{X_{L}=x,X_{H}=y, \Pi_{1}=L}]
\notag
\\
& \quad +\mathbf{P}[X=x\mid{X_{L}=x,X_{H}=y,\Pi_{1}=L,\Delta_{1}=R}]
\mathbf{P}[\Delta_{1}=R\mid{X_{L}=x,X_{H}=y, \Pi_{1}=L}].
\end{align}
The first probability on the right-hand side of equation (\ref{rhandfirone})
is equal to $1$, and the third probability is equal to $0$. Hence, equation \eqref{rhandfirone} simplifies to
\begin{equation}\label{rhandfironer}
\mathbf{P}[X=x\mid{X_{L}=x,X_{H}=y,\Pi_{1}=L}]
= \mathbf{P}[\Delta_{1}=A\mid{X_{L}=x,X_{H}=y, \Pi_{1}=L}].
\end{equation}
Similarly, we obtain the expression
\begin{equation}\label{rhandfirtwore}
\mathbf{P}[X=x\mid{X_{L}=x,X_{H}=y,\Pi_{1}=H}]
= \mathbf{P}[\Delta_{1}=R\mid{X_{L}=x,X_{H}=y, \Pi_{1}=H}]
\end{equation}
for the third probability
on the right-hand side of equation (\ref{rhandfir}). Using equations \eqref{rhandfironer} and \eqref{rhandfirtwore} on the right-hand of equation \eqref{rhandfir}, we have
\begin{align}\label{rhandfironerfinal-0}
&\mathbf{P}[X=x\mid{X_{L}=x,X_{H}=y}]
\notag
\\
&=\mathbf{P}[\Delta_{1}=A\mid{X_{L}=x,X_{H}=y, \Pi_{1}=L}]
\mathbf{P}[\Pi_{1}=L\mid{X_{L}=x,X_{H}=y}]
\notag
\\
&\quad + \mathbf{P}[\Delta_{1}=R\mid{X_{L}=x,X_{H}=y, \Pi_{1}=H}]
\mathbf{P}[\Pi_{1}=H\mid{X_{L}=x,X_{H}=y}]
\notag
\\
&=\mathbf{P}[\Delta_{1}=A\mid{X_{\Pi_{1}}=x,X_{\Pi_{2}}=y, \Pi_{1}=L}]
\mathbf{P}[\Pi_{1}=L\mid{X_{L}=x,X_{H}=y}]
\notag
\\
&\quad + \mathbf{P}[\Delta_{1}=R\mid{X_{\Pi_{2}}=x,X_{\Pi_{1}}=y, \Pi_{1}=H}]
\mathbf{P}[\Pi_{1}=H\mid{X_{L}=x,X_{H}=y}],
\end{align}
where $\Pi_{2}$ ($\neq \Pi_{1}$) is the salesperson who offers prices during the second time-period.

We now recall the assumption preceding Theorem \ref{th-1a} that tells us that the decision to accept or reject the first-come offer does not depend on who makes the offer -- the decision  depends only on the size of the offer. In probabilistic language, this means the equation
\[
\mathbf{P}[\Delta_{1}=\delta \mid{X_{\Pi_{1}}=x,X_{\Pi_{2}}=y, \Pi_{1}=\pi}]=\mathbf{P}[\Delta_{1}=\delta
\mid{X_{\Pi_{1}}=x,X_{\Pi_{2}}=y}]
\]
that must hold for every decision $\delta \in \{A,R\}$ (i.e., accept or reject) and for every salesperson $\pi\in\{L,H\}$, where $\Pi_{2}~(\neq \Pi_{1})$ denotes the salesperson who makes the offer during the second time-period. Hence, we have the equations
\begin{equation}\label{assumption-1a}
\mathbf{P}[\Delta_{1}=A\mid{X_{\Pi_{1}}=x,X_{\Pi_{2}}=y, \Pi_{1}=L}]=\mathbf{P}[\Delta_{1}=A
\mid{X_{\Pi_{1}}=x,X_{\Pi_{2}}=y}]
\end{equation}
and
\begin{equation}\label{assumption-1b}
\mathbf{P}[\Delta_{1}=R\mid{X_{\Pi_{2}}=x,X_{\Pi_{1}}=y, \Pi_{1}=H}]
=1-\mathbf{P}[\Delta_{1}=A
\mid{X_{\Pi_{1}}=y,X_{\Pi_{2}}=x}].
\end{equation}

\begin{note}
Given the description of our problem, we might naturally think that the decision variable $\Delta_{1}$ is independent of the hypothetical/speculative future value of $X_{\Pi_{2}}$, and thus the right-hand sides of equations (\ref{assumption-1a}) and (\ref{assumption-1b}) simplify by leaving out the second conditions associated with $X_{\Pi_{2}}$. We shall indeed consider this situation later, but at the moment we admit the possibility (cf.~Theorem \ref{th-1b}) that some clues about possible price offerings during the second time-period might be available to the buyer.
\end{note}

With the notation
\begin{equation}\label{strategy-s}
\mathcal{S}_w(v)=\mathbf{P}[\Delta_{1}=A
\mid{X_{\Pi_{1}}=v,X_{\Pi_{2}}=w}],
\end{equation}
the right-hand side of equation (\ref{assumption-1a}) is equal to $\mathcal{S}_y(x)$ and the right-hand side of equation (\ref{assumption-1b}) is equal to $1-\mathcal{S}_x(y)$. Consequently, equation (\ref{rhandfironerfinal-0}) turns into the following one
\begin{multline}\label{rhandfironerfinal}
\mathbf{P}[X=x\mid{X_{L}=x,X_{H}=y}]
\\
=\mathcal{S}_y(x)\mathbf{P}[\Pi_{1}=L\mid{X_{L}=x,X_{H}=y}]
+ (1-\mathcal{S}_x(y))
\mathbf{P}[\Pi_{1}=H\mid{X_{L}=x,X_{H}=y}].
\end{multline}
This is a desired expression for the first probability on the right-hand side of equation \eqref{expectation-first}. As to the second probability, analogous considerations lead to the equations
\begin{align}\label{rhandsecfinal}
&\mathbf{P}[X =y\mid{X_{L}=x,X_{H}=y}]
\notag
\\
&=\mathbf{P}[\Delta_{1}=A\mid{X_{L}=x,X_{H}=y, \Pi_{1}=H}]
\mathbf{P}[\Pi_{1}=H\mid{X_{L}=x,X_{H}=y}]
\notag
\\
&\quad + \mathbf{P}[\Delta_{1}=R\mid X_{L}=x,X_{H}=y, \Pi_{1}=L]
\mathbf{P}[\Pi_{1}=L\mid X_{L}=x,X_{H}=y]
\notag
\\
&=\mathcal{S}_x(y)\mathbf{P}[\Pi_{1}=H\mid{X_{L}=x,X_{H}=y}]
+ (1-\mathcal{S}_y(x))\mathbf{P}[\Pi_{1}=L\mid X_{L}=x,X_{H}=y].
\end{align}
Applying equations \eqref{rhandfironerfinal} and \eqref{rhandsecfinal} on the right-hand side of equation \eqref{expectation-first}, we have
\begin{align*}
&\mathbf{E} [X\mid{X_{L}=x,X_{H}=y}]
\\
&=x\Big (\mathcal{S}_y(x)
\mathbf{P}[\Pi_{1}=L\mid{X_{L}=x,X_{H}=y}]
+ (1-\mathcal{S}_x(y))
\mathbf{P}[\Pi_{1}=H\mid{X_{L}=x,X_{H}=y}] \Big )
\\
&\quad +y\Big (\mathcal{S}_x(y)
\mathbf{P}[\Pi_{1}=H\mid{X_{L}=x,X_{H}=y}]
+ (1-\mathcal{S}_y(x))
\mathbf{P}[\Pi_{1}=L\mid X_{L}=x,X_{H}=y] \Big )
\end{align*}
which becomes
\begin{multline}\label{expextation-third}
\mathbf{E} [X\mid{X_{L}=x,X_{H}=y}]
=x\Big (\mathcal{S}_y(x)p(x,y)
+ (1-\mathcal{S}_x(y))(1-p(x,y)) \Big )
\\
+y\Big (\mathcal{S}_x(y)(1-p(x,y))
+ (1-\mathcal{S}_y(x))p(x,y) \Big )
\end{multline}
with the probability
\[
p(x,y)=\mathbf{P}(\Pi_{1}=L\mid X_{L}=x,X_{H}=y)
\]
whose meaning was discussed before the formulation of Theorem \ref{th-1a}. We next rearrange the terms on the right-hand side of equation (\ref{expextation-third}) by separating the strategy-free and strategy-dependent terms:
\begin{multline}\label{expextation-3}
\mathbf{E} [X\mid{X_{L}=x,X_{H}=y}]
=x(1-p(x,y))+y p(x,y)
\\
+\mathcal{S}_y(x)\Big ( x p(x,y)- y p(x,y)\Big )
+\mathcal{S}_x(y)\Big ( y(1-p(x,y))- x(1-p(x,y))\Big ).
\end{multline}
Combining equations (\ref{expextation-3}) and (\ref{conditional-expectation-first}), we obtain the decomposition
\begin{equation}
\label{exp-1}
\mathbf{E} [X]=\mu_0+\mu_1(\mathcal{S}),
\end{equation}
where the strategy-free term is
\begin{align*}
\mu_0
&=\iint \Big ( x(1-p(x,y))+y p(x,y) \Big) dF_{X_L,X_H}(x,y)
\\
&= \mathbf{E}\big [ X_L(1-p(X_L,X_H))\big]
+\mathbf{E}\big [X_H p(X_L,X_H)\big]
\end{align*}
and the strategy-dependent term is
\begin{multline*}
\mu_1(\mathcal{S})
=\iint \mathcal{S}_y(x)\Big ( x p(x,y)- y p(x,y)\Big )dF_{X_L,X_H}(x,y)
\\
+\iint \mathcal{S}_x(y)\Big ( y(1-p(x,y))- x(1-p(x,y))\Big )
 dF_{X_L,X_H}(x,y).
\end{multline*}
Rewriting the latter equation in terms of the joint density $f_{X_L,X_H}(x,y)$, and also slightly changing some notation, we arrive at the equation
\begin{equation}\label{int-0}
\mu_1(\mathcal{S})=\iint \mathcal{S}_w(v) h(v,w) dvdw,
\end{equation}
where $h(v,w)$ is defined by equation (\ref{h-vw}). Since $\mathcal{S}_w(v)$ is a probability and can therefore take values only in the unit interval $[0,1]$, integral (\ref{int-0}) achieves its minimal value when
\[
\mathcal{S}_w(v)=
\left\{
  \begin{array}{ll}
    1 & \hbox{ if}\quad h(v,w)\le 0,\\
    0 & \hbox{ if}\quad h(v,w)> 0.
  \end{array}
\right.
\]
This completes the proof of Theorem \ref{th-1b}.

We next deal with the case (cf.\, Theorem \ref{th-1a}) when the decision random variable $\Delta_{1}$ is independent of $X_{\Pi_{2}}$ and thus  $\mathcal{S}_w(v)$ does not depend on the price $w$ offered during the second time-period. Hence, instead of $\mathcal{S}_w(v)$, we now deal with the strategy function
\begin{equation}
\label{strategy-s2}
\mathcal{S}(v)=\mathbf{P}[\Delta_{1}=A \mid{X_{\Pi_{1}}=v}].
\end{equation}
Consequently, the above defined $\mu_1(\mathcal{S})$ reduces to the integral
\begin{equation}\label{int-1}
\mu_1(\mathcal{S})=\int \mathcal{S}(v)h(v) dv,
\end{equation}
where $h(v)$ is defined by equation (\ref{h-v}). Integral (\ref{int-1}) achieves its minimal value when
\[
\mathcal{S}(v)=
\left\{
  \begin{array}{ll}
    1 & \hbox{ if}\quad h(v)\le 0,\\
    0 & \hbox{ if}\quad h(v)> 0.
  \end{array}
\right.
\]
This completes the proof of Theorem \ref{th-1a}.

\section*{Acknowledgements}

We are indebted to the Editor and an anonymous referee for constructive criticism, queries, and suggestions that have greatly influenced our work on the revision. We are also grateful to Mart\'{i}n Egozcue for his much appreciated advice and generosity. The research of both authors has been supported by the grant ``From Data to Integrated Risk Management and Smart Living: Mathematical Modelling, Statistical Inference, and Decision Making'' (2016--2021) awarded to the second author by the Natural Sciences and Engineering Research Council of Canada.

%
%
%
%

\end{document}